\documentclass[aps,pre,twocolumn,showpacs]{revtex4-1}

\usepackage{amssymb}
\usepackage{amsmath}
\usepackage{amsthm}
\usepackage{graphicx}
\usepackage{amsfonts}
\usepackage{color}
\usepackage{times}
\usepackage{natbib}
\usepackage{soul}
\usepackage{booktabs}
\usepackage{pifont}

\usepackage{hyperref}
\hypersetup{colorlinks=true}
\usepackage{mathpazo}

\RequirePackage{fix-cm}

\usepackage[normalem]{ulem}

\newcommand{\myskip}[1]{}   
\newcommand{\BEQ}{\begin{eqnarray}}      
\newcommand{\EEQ}{\end{eqnarray}}      
\newcommand{\BEA}{\begin{eqnarray}}      
\newcommand{\EEA}{\end{eqnarray}}

\newcommand{\tr}{{\rm tr}}

\newcommand{\D}{\hat{\mathcal{D}}}
\newcommand{\R}{\hat{\mathcal{R}}}

\newcommand{\Ruu}{\hat{R}_{\uparrow \uparrow}}
\newcommand{\Rud}{\hat{R}_{\uparrow \downarrow}}
\newcommand{\Rdu}{\hat{R}_{\downarrow \uparrow}}
\newcommand{\Rdd}{\hat{R}_{\downarrow \downarrow}}

\newcommand{\ketup}{|\hspace{-1mm}\uparrow\rangle}
\newcommand{\ketdown}{|\hspace{-1mm}\downarrow\rangle}
\newcommand{\braup}{\langle\uparrow\hspace{-1mm}|}
\newcommand{\bradown}{\langle\downarrow\hspace{-1mm}|}

\usepackage{graphicx}
\usepackage{dcolumn}
\usepackage{bm}
\usepackage{mathtools}
\usepackage{lipsum}

\renewcommand{\thesection}{\arabic{section}}

\renewcommand{\thesection}{\arabic{section}}

\begin{document}

\title{Dynamics of quantum measurements employing two Curie-Weiss apparatuses}

\author{Mart{\'i} Perarnau-Llobet}
\email{marti.perarnau@mpq.mpg.de} 
\affiliation{Max-Planck-Institut f\"ur Quantenoptik, Hans-Kopfermann-Str. 1, D-85748 Garching, Germany}
\affiliation{ICFO-Institut de Ciencies Fotoniques, The Barcelona Institute of Science and Technology, 08860 Castelldefels, Barcelona, Spain}

\author{Theodorus Maria Nieuwenhuizen}
\affiliation{Institute for Theoretical Physics, University of Amsterdam, Science Park 904, 1090 GL  Amsterdam, The Netherlands} 
\affiliation{International Institute of Physics, UFRG,  Anel Vi\'ario da UFRN - Lagoa Nova, Natal - RN, 59064-741, Brazil}

\begin{abstract}
Two types of quantum measurements, measuring the spins of an entangled pair and attempting to measure a spin at either of two positions,
are analysed dynamically by apparatuses of the Curie-Weiss type. The outcomes comply with the standard postulates. 
\end{abstract}
 
 

\maketitle

 \setcounter{section}{0}
 \renewcommand{\thesection}{\arabic{section}}
 \section*{ Introduction}
 \setcounter{equation}{0} \setcounter{figure}{0}
 \renewcommand{\thesection}{\arabic{section}.}

Describing quantum measurements as a joint evolution between  the tested system (S) and the apparatus (A) is an aim that goes back to the early days 
of quantum mechanics \cite{vonneumann,wh,{deMuynck},other3}. Far from being solved, this quest is still an active area of research, ranging from specific models  to general 
mechanisms to describe the measurement process. Recently, the literature on this subject has been extensively reviewed in Ref. \cite{OpusABN}, while recent foundational studies
in quantum physics can be found in, e.g., Refs \cite{AI,AII,AIII}.  

In this article we focus our attention to the Curie-Weiss (CW) model for a spin measurement \cite{ABNcw2003}. 
It allows a dynamical description of a projective measurement of a spin system using a magnetic memory, which takes the role of the apparatus. 
The probabilities of the different outcomes of the measurement are then derived through the common evolution of the spin and the magnetic memory. 
The interpretation of the final density matrix and its connection to the {\it measurement problem} \cite{{deMuynck},other1} was recently discussed in Refs. \cite{OpusABN,OpusII}, this is why here we shall stick to the standard interpretation.

The present paper deals with several questions involving two CW apparatuses. First, we apply  the CW model to the measurement of a spin projection in an entangled state, and then consider an attempt to measure a spin projection simultaneously by two spatially 
separated apparatuses.  The results obtained through the dynamical equations will turn out to be in agreement with the standard measurement postulates, 
giving further support for the CW-model as a proper model for a projective measurement.
The intricate situation where two non-commuting variables are measured 
simultaneously, so that the apparatuses influence each other through their couplings to the tested system, was considered by us in Ref. \cite{marti}.

In section 1 we recall the essentials of the CW model. In section 2 we apply it to measure a spin projection of an entangled pair, or measure both partners with two
CW apparatuses. In section 3 we consider a simultaneous measurement of a spin projection by  two well separated apparatuses.
We close with a summary.

 \setcounter{section}{0}
 \renewcommand{\thesection}{\arabic{section}}
\section{The Curie-Weiss model for a quantum measurement process} 
 \setcounter{equation}{0} \setcounter{figure}{0}
 \renewcommand{\thesection}{\arabic{section}.}
\label{SecCW}

We start by giving a short explanation of the Curie-Weiss (CW) model for a measurement process.
It was introduced in \cite{ABNcw2003}; for a detailed description, we refer the reader to \cite{OpusABN}.
This model describes the measurement of the $z$-component of a spin $\frac{1}{2}$-system by an apparatus in the form of a magnetic memory, 
which starts out as a paramagnet, but will end up in an up-or-down magnetised state (Ising magnet). 
Due to the macroscopic size of the apparatus, the purely unitary evolution of $SA$ can describe a measurement process for the relevant degrees of freedom, i.e., the establishment of macroscopic robust correlations between $S$ and $A$.
The post-measurement state of the spin (up-or-down along the $z$-axis) is inferred from the sign of the magnetisation (ideal measurement).

A crucial property of any measurement is the pointer variable, that is, the physical property of $A$ from which one can infer information about $S$. 
In the CW-model, this role is played by the magnetisation of $A$. Indeed, $A$ is represented by a magnet $M$, a collection of $N$ spins 
$\hat{\sigma}^{(n)}_a$ ($a=x,y,z$), in weak thermal contact with a bath $B$. The operator for the magnetisation of $M$ is given by,

\begin{align}
\hat{m}=\frac{1}{N} \sum_{n=1}^N \hat{\sigma}_z^{(n)}.
\end{align}
and $m$ describes the values it can take, viz. $-1\le m\le 1$.
This pointer variable of A couples to the measured variable of S, $\hat{s}_z$, as the product
\begin{equation}
 H_{\rm SA}=H_{SM}=-Ng\hat{m} \otimes \hat{s}_z, 
 \label{HSA}
\end{equation}
with the factor $N$ for convenience, and the coupling $g$ turned on at the beginning  of the measurement and switched off at the end.

More specifically, the total Hamiltonian of $SA$ can be decomposed as,
\begin{align}
H= H_S+H_{SA} + H_A=H_{SM}+H_M + H_{MB}+H_B,
 \label{recapHtot}
\end{align}
where $H_S= 0$ indicates that the system does not evolve during the measurement, a condition for its ideality.
The self-Hamiltonian of the magnet reads
 \begin{align}
H_{\rm M}=-NJ_2\frac{\hat{m}^2}{2}-NJ_4\frac{\hat{m}^4}{4}.
\end{align}
with $J_2\ge0$ and $J_4>0$.
Furthermore, B is a bosonic bath,
\begin{equation}
	\hat{H}_{{\rm B}}  =\sum_{n=1}^{N}\sum_{a=x,y,z}\sum_{k}\hbar\omega_{k}\hat{b}_{k,a}^{\dagger\left(  n\right)  }\hat{b}_{k,a}^{\left(  n\right)  }
	\label{H_{B}}
\end{equation}
where $\hat{b}_{k,a}^{\dagger\left(  n\right)  }$ are phonon modes with eigenfrequencies $\omega_k$, which couple independently to each of the $N$ spins of $M$,
\begin{equation}
	\hat{H}_{{\rm M}{\rm B}}  =\sqrt{\gamma}\sum_{n=1}^{N}\sum_{a=x,y,z}\hat{\sigma}_{a}^{\left(  n\right)  }\hat{B}_{a}^{\left(n\right)  }
	\label{couplingBM}
\end{equation}
where $\hat{B}_{a}^{\left(n\right)} $ are phonon operators given by:
\begin{equation}
		\label{Ba}
		 \hat{B}_{a}^{\left(n\right)  }=\sum_{k}\sqrt{c\left(  \omega_{k}\right)  }\left(  \hat{b}_{k,a}^{\left(  n\right)  }+\hat{b}_{k,a}^{\dagger\left(  n\right)  }\right).
\end{equation}
Summarising, $A$ is composed of a magnet $M$ in weak contact with a bosonic bath, and is coupled to $S$ through \eqref{HSA}.

Initially, $SMB$ are uncoupled and described by a tensor product of density matrices,
\begin{align}
\D(0)= \hat{r}(0) \otimes \hat{R}_M(0) \otimes \hat{R}_B(0) 
\end{align}
where $\hat{r}(0)$ is the arbitrary spin state,  $\hat{R}_M(0)$ is a paramagnetic state of the magnet,
\begin{equation}
\hat{R}_M= \frac{1}{2^N} \bigotimes_{n=1}^N \mathbb{I}^{(n)}
\label{RMinitial}
\end{equation}
expressing that each spin component has probability $\frac{1}{2}$ to be up or down along any chosen axis. This is state is so that the pointer is initialized at zero, $\tr \left( \hat{R}_M \hat{m}\right)=0$. Other possible initial states of $M$ are discussed in \cite{OpusABN,marti}.
Lastly, the factor  $\hat{R}_B(0)$ is a thermal state of the bath,
\begin{align}
\hat{R}_B(0) = \frac{e^{-\beta \hat{H}_B}}{\mathcal{Z}_B}.
\end{align} 
The temperature $1/\beta$ is chosen low enough such that the paramagnetic ($m=0$) state is metastable.
For $J_4>0$ large enough the magnet may undergo a first order phase transition from the metastable state with $m=0$ 
 to either of its stable ferromagnetic states, with magnetisation up or down, $m\to \pm m_{\rm F}$. 
 The  transition will be triggered during the measurement process.
The final magnetisation $\pm Nm_{\rm F}$  can be read off, which is the registration of the measurement.

Now, the evolution of $\D(t)$ is dictated by the Liouville-von Neumann equation of motion, 
\begin{equation}
 i\hbar \frac{{d} \D}{{d}t}=[H,\D].
 \label{recapeqmotion}
\end{equation}
The relevant degrees of freedom are those of the state of $SM$, i.e., the measured system and the pointer. Hence, one is in fact interested in the evolution of the reduced state of $SM$,
\begin{align}
\hat{D}(t)=\tr_B \D(t)
\end{align}
where $\D(t)$ is the state of $SMB$. By tracing out the bath, one can obtain a dissipative equation for the evolution of $SM$ by using standard open systems techniques owing to the weak coupling between $M$ and $B$. One then obtains the following master equation for the evolution of $SM$ under its Hamiltonian $\hat H_0=\hat H_{\rm SM}+\hat H_{\rm M}$ \cite{OpusABN},
\begin{align}
\frac{d\hat{D}}{dt}-\frac{1}{i\hbar}[\hat{H}_0,\hat{D}]=&\frac{\gamma}{\hbar^2} \int_0^t du \sum_{n,a} \bigg(K(u) [\hat{\sigma}_a^{(n)}(u)\hat{D},\hat{\sigma}^{(n)}_a]
\nonumber\\
&+K(-u)[\hat{\sigma}^{(n)}_a,\hat{D}\hat{\sigma}_a^{(n)}(u)]\bigg)+\mathcal{O}(\gamma^2)
\label{MasterEq}
\end{align}
where $K(u)$ is the correlation function of the quasi-Ohmic bath, see ref. \cite{OpusABN},  and
\begin{align}
\hat{\sigma}_a^{(n)}(t)= e^{i\hat H_0t/\hbar}\hat{\sigma}_a^{(n)}e^{-i\hat H_0t/\hbar}
\end{align}
is the time-evolution of $\hat{\sigma}_a^{(n)}$ under $\hat H_0$.
It is now convenient to expand $\hat{D}(t)$ using the basis of the tested spin,
\begin{align}
 \hat{D}(t)&=\ketup \braup \Ruu(t) +\ketdown \bradown \Rdd(t)
 \nonumber\\ 
 &\, +\ketup \bradown \Rud(t)+\ketdown \braup \Rdu (t).
\label{recapgeneralstate}
\end{align}
This allows us to decompose the master equation \eqref{MasterEq} into four equations of motion, which happen to be   independent
in this model for ideal projective measurements,
\begin{align}
\frac{d\hat{R}_{ij}}{dt}-&\frac{\hat{H}_i\hat{R}_{ij}-\hat{R}_{ij}\hat{H}_j}{i\hbar}=\nonumber\\
&=\frac{\gamma}{\hbar^2} \int_0^t du \sum_{n,a} \bigg(K(u) [\hat{\sigma}_a^{(n)}(u)\hat{R}_{ij},\hat{\sigma}^{(n)}_a]
\nonumber\\
&+K(-u)[\hat{\sigma}^{(n)}_a,\hat{R}_{ij}\hat{\sigma}_a^{(n)}(u)]\bigg)+\mathcal{O}(\gamma^2)
 \label{recapeqRij}
\end{align}
where 
$\hat{H}_{i}=-gNs_{i}\hat{m}-\frac{1}{2}NJ_2\hat m^2-\frac{1}{4}NJ_4\hat m^4$ and $s_i=\pm 1$ for up or down, respectively.
Explicitly, we obtain for the diagonal elements,
\begin{align}
i\hbar\frac{d\hat{R}_{\uparrow \uparrow}}{dt}=f(\hat{R}_{\uparrow \uparrow})
\label{recapvdiag}
\end{align}
where $f(\hat{R}_{\uparrow \uparrow})$ stands for the right hand side of \eqref{recapeqRij} for the  $\uparrow \uparrow$ case, 
and similarly for $\hat{R}_{\downarrow \downarrow}$. For the off-diagonal elements we obtain
\begin{align}
 i\hbar\frac{d\hat{R}_{\uparrow \downarrow}}{dt}+2Ng\hat{m} \hat{R}_{\uparrow \downarrow}=f(\hat{R}_{\uparrow \downarrow})
\label{miseqoffdiag}
\end{align}
In \cite{OpusABN}, the equations of motion \eqref{recapeqRij} are solved.
 The off-diagonal terms  $\Rud(t)$ and $\Rdu(t)$ first gain phases due to the interaction $g$, which at the level of $S$ creates a dephasing process. Later, in a time scale governed by $1/\gamma$, they disappear due to a decoherence process induced by the thermal bath.
The diagonal terms $\Ruu(t)$, $\Rdd (t)$ reach the pointer states due to the combined interaction of $M$ with $S$ and $B$. 
These pointer states are the up-or-down  ferromagnetic states $\hat{R}_{\Uparrow}$, $\hat{R}_{\Downarrow}$, 
for which $m\approx \pm m_{\rm F}$ with $m_{\rm F}$ close to but not equal to $1$, and associated 
respectively with an outcome $\pm \frac{\hbar}{2}$ of the measurement of $\hat s_z$. 
In conclusion, the final state of $S$+$M$ indeed takes the form expected from the postulates \cite{OpusABN},
\begin{eqnarray}
 \label{recapDtf}
\hspace{-5mm} \hat{D}(t_{\rm f})=r_{\uparrow\uparrow}(0)|\hspace{-1mm}\uparrow\rangle\langle\uparrow\hspace{-1mm}|\otimes{\hat{ R}}_{\Uparrow}+
 r_{\downarrow\downarrow}(0)|\hspace{-1mm}\downarrow\rangle\langle\downarrow\hspace{-1mm}|\otimes{\hat{R}}_{\Downarrow} .
\end{eqnarray}
For the purposes of this work, it is enough to have the form of the equations \eqref{recapeqRij} and that of the final state \eqref{recapDtf}.

 \renewcommand{\thesection}{\arabic{section}}
\section{Measuring a spin of an entangled pair}
 \setcounter{equation}{0} \setcounter{figure}{0}
 \renewcommand{\thesection}{\arabic{section}.}
\label{entangledstate}

In this section we first discuss the CW-model in the context of a measurement of a spin $z$-projection of a maximally entangled state. 
Next we go on to measure both spins.

Let $a$, $b$ denote each of the two spins.  Consider then an EPR state,
\begin{equation}
|\psi \rangle= \frac{1}{\sqrt{2}} \left(|\hspace{-1mm}\uparrow_a \downarrow_b \rangle + |\hspace{-1mm}\downarrow_a \uparrow_b \rangle \right)
\label{initialState}
\end{equation}
which corresponds to the density matrix
\begin{align}
\hat{r}=&\frac{1}{2} \big(\ketup \braup_a \otimes \ketdown \bradown_b+\ketup \bradown_a \otimes \ketdown \braup_b 
\nonumber\\
&+  \ketdown \braup_a  \otimes \ketup \bradown_b +  \ketdown \bradown_a \otimes \ketup \braup_b \big) .
\end{align}
Imagine now that the two spins are spatially separated, so that the apparatus only interacts with one of them, say $a$. 
The Hilbert space is: $\mathcal{H}_{S_a}  \otimes \mathcal{H}_{A}\otimes \mathcal{H}_{S_b}$. The total Hamiltonian  reads,
\begin{align}
\hat{H}_T&=H\otimes \mathbb{I}_{S_b}=(\hat{H}_{\rm M}+\hat{H}_{\rm B}+\hat{H}_{\rm MB}+\hat{H}_{\rm S_a A})\otimes \mathbb{I}_{S_b}.
\end{align}
Let us stress here that the Hamiltonian $\hat{H}_{\rm MB}=-Ng\hat{m} \otimes \hat{s}_z^{(a)}$ is responsible for selecting the preferred basis $z$ of the measurement.

When tracing out B, the reduced density matrix of the system at any time $t$  will have the form (since $\hat{H}_T$ commutes with $\hat{s}_z$ of both particles):
\begin{align}
	&\hat{D}(t)=\frac{1}{2} (\ketup \braup_a \otimes R_{\uparrow \uparrow}(t) \otimes  \ketdown \bradown_b   \nonumber\\
	+& \ketup \bradown_a \otimes R_{\uparrow \downarrow}(t) \otimes   \ketdown \braup_b+
	\ketdown \braup_a \otimes R_{\downarrow \uparrow}(t) \otimes    \ketup \bradown_b
	\nonumber\\
	+&\ketdown \bradown_a \otimes R_{\downarrow \downarrow}(t) \otimes   \ketup \braup_b)
\end{align}
where, $\hat{R}_{ij}$ evolve independently following equations (\ref{recapeqRij}). Then, we can directly apply the results of \cite{OpusABN}, which imply that $R_{\uparrow \downarrow}(t)$ and $R_{\downarrow \uparrow}(t)$ decay to zero due to the action of $B$; furthermore,  $R_{\uparrow \uparrow}(t)$ and $R_{\downarrow \downarrow}(t)$ tend to $R_{\Uparrow}$ and $R_{\Downarrow}$ respectively. Hence, the final state takes the form,
\begin{align} \label{twospin-oneapp}
 \hat{D}(t_f)&=\frac{1}{2} \ketup \braup_a \otimes R_{\Uparrow} \otimes \ketdown \bradown_b \nonumber\\
 &+\frac{1}{2} \ketdown \bradown_a \otimes R_{\Downarrow} \otimes \ketup \braup_b .
\end{align}
From this state, we see that by reading the pointer of A we can infer the state of the (measured) spin $a$ as well as the state of spin $b$, 
which did not interact with A. This is exactly the same behaviour that one would have obtained by applying the standard projective measurement postulates 
to \eqref{initialState}. In this sense, we straightforwardly see that the CW-model also provides the expected results by quantum mechanics when considering projective 
measurements of an entangled state.

\subsection{Adding a second Curie-Weiss apparatus}

To finalise this section, consider the role of a second, similar apparatus A' measuring the $z$ component of a spin $b$ by an interacting Hamiltonian of the form \eqref{recapHtot} with $\hat{H}_{\rm MB'}=-Ng\hat{m}' \otimes \hat{s}_z^{(b)}$. The needed steps are fully similar.
As expected, such an apparatus would only confirm the obtained result by the first apparatus (note that both apparatuses are measuring the same spin component $z$),  in such a way that the final state would read,
\begin{align} \label{twospin-twoapp}
 \hat{D}(t_f)=&\frac{1}{2} \bigg( \ketup \braup_a \otimes R_{\Uparrow} \otimes \ketdown \bradown_b \otimes R_{\Downarrow}' \nonumber\\
 & + \ketdown \bradown_a \otimes R_{\Downarrow} \otimes \ketup \braup_b \otimes R_{\Uparrow}' \bigg) .
\end{align}
For example, when reading off the apparatus of $s_a$,  see (\ref{twospin-oneapp}), and finding it up (the first term), one would expect that 
if the second apparatus $A'$ is present, it would produce a result down, and this is indeed expressed in the first term of (\ref{twospin-twoapp}).

In popular terms this can be phrased as: Because the state is entangled, Alice can predict from her measurement what Bob will find in his.
This is in no way the result of ``faster than light communication'' but merely an expression of the (quantum) correlations.

 \renewcommand{\thesection}{\arabic{section}}
\section{Trying to measure a spin simultaneously by two locally separated detectors}
 \setcounter{equation}{0} \setcounter{figure}{0}
 \renewcommand{\thesection}{\arabic{section}.}

In this section we consider a setting where, in principle,  $S$ can interact simultaneously with two apparatuses 
-- both of them attempting to measure $\hat{s}_z$ -- located at two spatially separated positions. 
It is then clear that we need to account for the spatial density matrix $\hat \rho$ of S.

\subsection*{One detector confined in space}

We start by considering a simpler scenario, namely one  apparatus located in a region of  a one-dimensional space $\mathcal{R}=(a,b)$. This can be expressed straightforwardly by changing the Hamiltonian (\ref{recapHtot}) to,
\begin{equation}
 H_T=H_{A}+V(\hat{r}) \otimes H_{SA}
 \label{HTR}
\end{equation}
with
\begin{displaymath}
   V(x) = \left\{
     \begin{array}{lr}
       -k \hspace{7mm}  x \in \mathcal{R}=(a,b)\\
       0 \hspace{10mm}  x \in \mathcal{\overline{R}}
     \end{array}
   \right.
\end{displaymath} 
where $\overline{\mathcal{R}}$ is the union of the regions $(-\infty,a)$ and $(b,\infty)$.  Clearly, the measurement can only happen when the spin is in the interval $(a,b)$, that is, 
when it is close enough to $A$. The initial state of the whole system is then
\begin{align}
 \D_T(0)=&\hat{\rho}(0)\otimes \hat{r}(0)\otimes \hat{R}_{\rm M} (0) \otimes \hat{R}_{\rm B} (0)
 \nonumber\\ 
 =&\int \int dx dy \hspace{1mm} \rho(x,y;0) |x\rangle \langle y|  \hat{r}(0) \hat{R}_{\rm M} (0)  \hat{R}_{\rm B} (0),
\end{align}
where $\rho(x,y;0)$ is the spatial part of the density matrix and the other factors are as before.
Since the spatial part of $H_T$ is a piecewise constant in the various regions of space, we can take for $\D_T (t)$ the form:
\begin{align}
\D_T(t)&= \int_{\mathcal{R}} \int_{\mathcal{R}} dx dy \hspace{1mm} \rho(x,y) |x\rangle \langle y| \otimes \D_I (t)
\nonumber\\ 
&+ \int_{\overline{\mathcal{R}}} \int_{\overline{\mathcal{R}}} dx dy \hspace{1mm} \rho(x,y) |x\rangle \langle y| \otimes \D_{II} (t)
\nonumber\\
&+ \int_{\mathcal{R}} dx \int_{\overline{\mathcal{R}}} dy  \hspace{1mm} \rho(x,y) |x\rangle \langle y| \otimes \D_{III} (t) \nonumber \\
&+ \int_{\overline{\mathcal{R}}} dx \int_{\mathcal{R}}  dy  \hspace{1mm} \rho(x,y) |x\rangle \langle y| \otimes \D_{III} ^\dagger(t) .
\label{expDansatz}
\end{align}
Assuming no other spatial aspects and a constant flux of particles, $\rho(x,y;t)=\rho(x,y;0)$ is time-independent and denoted as $\rho(x,y)$. 
Now each $\D_i (t)$ (with $i=I,II,III$) will evolve independently. Indeed, if we insert this Ansatz into the equation of motion, 
\begin{equation}
	i \hbar \frac{d \D_T}{dt}=[H_T,\D]=[H_{A},\D]+[V(\hat{r}) H_{SA},\D],
\end{equation}
we reach for the right hand side,


\begin{align}
&\int_{\mathcal{R}} \int_{\mathcal{R}} dx dy \hspace{1mm} \rho(x,y) |x\rangle \langle y| [H_{A}-k H_{SA},\D_I ] 
\nonumber\\
&+ \int_{\overline{\mathcal{R}}} \int_{\overline{\mathcal{R}}} dx dy \hspace{1mm} \rho(x,y) |x\rangle \langle y| [H_{A},\D_{II}]
\\ & +\int_{\mathcal{R}} dx \int_{\overline{\mathcal{R}}} dy \hspace{1mm} \rho(x,y) |x\rangle \langle y|  \{[H_{A},\D_{III}]-kH_{SA}\D_{III}\} \nonumber \\ 
&+ h.c.,
\end{align}
where in the last term we took account of the fact that  in its integration intervals $V(x)=-k$ and $V(y)=0$. 
This immediately leads to an uncoupled equation of motion for each $\D_i (t)$,
\begin{align}
&i \hbar \frac{d \D_I}{dt}=[H_{A},\D_I]-k[H_{SA},\D_I],\nonumber\\
&i \hbar \frac{d \D_{II}}{dt}=[H_{A},\D_{II}],\nonumber\\
&i \hbar \frac{d \D_{III}}{dt}=[H_{A},\D_{III}]-k H_{SA}\D_{III}
\end{align}
The first equation of motion accounts for the spin located in the region $\mathcal{R}$. Comparing it to (\ref{recapeqmotion}), we see that we have exactly the same situation except from the factor $-k$, which comes from the potential and can be absorbed in its strength $g$ (without relevant changes for the process). 
Therefore, as expected, in this region the measurement can take place. The second equation of motion accounts for the spin located outside $\mathcal{R}$. In this region nothing happens: there is no interaction between S and A. In this case, we expect A to remain in the initial state. 
(It will not concern us that on a much longer scale time than that of the measurement, A will reach thermodynamic equilibrium because of its interaction with $B$.)

More interesting is the third equation of motion. It accounts for the spatial coherent terms of the density matrix. 
We obtain the following equations of motion for its components $\hat{R}_{ij}$ defined by (\ref{recapgeneralstate}):
\begin{eqnarray}
i\hbar\frac{d\hat{R}_{\uparrow \downarrow}}{dt}-kgN\hat{m}\hat{R}_{\uparrow \downarrow}=f(\hat{R}_{\uparrow \downarrow}),\\
i\hbar\frac{d\hat{R}_{\uparrow \uparrow}}{dt}-kgN\hat{m}\hat{R}_{\uparrow \uparrow}=f(\hat{R}_{\uparrow \uparrow})
\label{miseqmotionRij}
\end{eqnarray}
and similarly for $\hat{R}_{\uparrow \downarrow}$ and $\hat{R}_{\downarrow \downarrow}$. Regarding the off diagonal terms we have the same equations of motion as   (\ref{miseqoffdiag}) if we replace $-kg \rightarrow 2g$. Therefore, those terms decay due to a dephasing plus decoherence effect \cite{OpusABN}. On the other hand, the diagonal terms have a different time evolution from (\ref{recapvdiag}), and in fact  follow exactly the same evolution as $\hat{R}_{\uparrow \downarrow}$, hence also disappearing. 

This consideration confirms that the interference terms of the density matrix can not be observed by the pointer, and they die out because of the effect of the spatially confined apparatuses.  Putting everything together, we can write the remaining terms of the final state as, 
\begin{align}
D_T(t_f)=& \int_{\mathcal{R}} \int_{\mathcal{R}} dx dy \hspace{1mm} \rho(x,y) |x\rangle \langle y| \otimes \hat{D} (t_f)
\nonumber\\
& 
+\int_{\overline{\mathcal{R}}} \int_{\overline{\mathcal{R}}} dx dy \hspace{1mm} \rho(x,y) |x\rangle \langle y| \otimes \hat{D} (0)
\end{align}
with $ \hat{D} (t_f)$ given in (\ref{recapDtf}) and $ \hat{D} (0)$ just given by $\hat{r} (0)\otimes \hat{R}_M (0)$. At $t_f$, the spin is either in $\mathcal{R}$ 
or in $\overline{\mathcal{R}}$: If the apparatus clicks, then the spin $\mathit{is}$ there.

\subsection*{Two detectors}

Now it is straightforward to extend the considerations to the scenario where two independent detectors, which are spatially separated, act on the system. The detectors are located in  $\mathcal{R}_1$ and $\mathcal{R}_2$, with $\mathcal{R}_1 \cap \mathcal{R}_2={\O}$ and $\mathcal{R}_3 \equiv {\mathbb{R}- \mathcal{R}_1\cup \mathcal{R}_2}$.  The Hilbert space is $\mathcal{H}_{position}\otimes \mathcal{H}_{spin} \otimes \mathcal{H}_{A_1} \otimes \mathcal{H}_{A_2}$ with $ \mathcal{H}_{A_1}$ and $ \mathcal{H}_{A_2}$ the Hilbert space of the apparatuses located in $\mathcal{R}_1$ and $\mathcal{R}_2$ respectively. Then, following \eqref{HTR}, we take $H$ as,
\begin{equation}
 H_T=H^{(1)}_{A}+H^{(2)}_{A}
 +V_1(\hat{r}) \otimes H^{(1)}_{SA}
 +V_2(\hat{r}) \otimes H^{(2)}_{SA},
\end{equation} 
with


\begin{displaymath}
   V_i(x) = \left\{  \begin{array}{lr}       -k \hspace{7mm}  x \in \mathcal{R}_i  \\
       0 \hspace{10mm}  x \in  \overline{\mathcal{R}}_i
     \end{array}   \right. \qquad (i=1,2)
     \label{inttwoapparatuses} \end{displaymath}

The initial state of the system is:
\begin{align}
 \D_T(0)&=\hat{\rho}\otimes \hat{r}(0)\otimes \R_{A_1}(0) \otimes \R_{A_2}(0)
\nonumber\\ 
 &=\hat{\rho}\otimes \hat{r}(0)\otimes \hat{R}^{(1)}_{M} (0) \otimes \hat{R}^{(1)}_{B} (0)\otimes \hat{R}^{(2)}_{M} (0) \otimes \hat{R}^{(2)}_{B} (0)
 \end{align}
We can expand the total state $\D_T (t)$ of $S+A_1+A_2$,  as we did in (\ref{expDansatz}), into 9 independent terms (coming from the 9 possible domains 
of integration ($\mathcal{R}_1$, $\mathcal{R}_2$ and $\mathcal{R}_3$) of the double space integrals). From these 9 terms, 3 terms correspond to well 
a localized density matrix in a region $\mathcal{R}_i$ ($i=1,2,3$) of space. Four other terms correspond to combinations of  $\mathcal{R}_3$ with 
$\mathcal{R}_1$ or $\mathcal{R}_2$. All these 7 terms can be worked out as in the previous section. The remaining two terms contain interference terms 
that involve the density matrix with the two detectors. They have the form,

\begin{equation}
 \int_{\mathcal{R}_1} \int_{\mathcal{R}_2} dx dy \hspace{1mm} \rho(x,y) |x\rangle \langle y| \otimes \D_{int} (t)
\end{equation}
where $ \D_{int} (t)$ has the form
\begin{align}
\D_{int} (t)&= \ketup \braup \otimes \R^{(1)}_{\uparrow\uparrow}\otimes \R^{(2)}_{\uparrow\uparrow}  +\ketup \bradown \otimes
\R^{(1)}_{\uparrow\downarrow}\otimes \R^{(2)}_{\uparrow\downarrow} 
\nonumber\\
&
+\ketdown \braup \otimes \R^{(1)}_{\downarrow\uparrow}\otimes \R^{(2)}_{\downarrow\uparrow} 
+ \ketdown \bradown \otimes \R^{(1)}_{\downarrow\downarrow}\otimes \R^{(2)}_{\downarrow\downarrow} 
\end{align}
Tracing out the bath and inserting $\D_{int} (t)$ in the equation of motion we reach the following equations of motion for the reduced matrices $\hat{R}_{ij}=\tr_{\rm B} \R_{ij}$,
\begin{align}
	&i\hbar\frac{d\hat{R}^{(1)}_{ij}\otimes \hat{R}^{(2)}_{ij}}{dt}-kNg\hat{m}^{(1)}\hat{R}^{(1)}_{ij}\otimes \hat{R}^{(2)}_{ij}-
	\nonumber\\ 
	&kNg\hat{m}^{(2)}\hat{R}^{(1)}_{ij}\otimes \hat{R}^{(2)}_{ij}=
	f(\hat{R}^{(1)}_{ij})\otimes \R^{(2)}_{ij}
+\R^{(1)}_{ij}\otimes f(\hat{R}^{(2)}_{ij})
\end{align}
which leads to an equation for each component,
\begin{equation}
	i\hbar\frac{d\hat{R}^{(k)}_{ij}}{dt}-kNg\hat{m}\hat{R}^{(k)}_{ij}=f(\hat{R}_{ij}^{(k)}), 
\end{equation}
with $k=1,2$ 
for each detector. Those equations have the same form as (\ref{miseqmotionRij}),  so we expect the interference terms between the two apparatuses to be physically unobservable and to tend to zero due to the action of the magnets and their baths \cite{OpusABN}. 

Putting everything together, we keep for the final state
\begin{align}
 D_T(t_f)&= \int_{\mathcal{R}_1} \int_{\mathcal{R}_1} dx dy \hspace{1mm} \rho(x,y) |x\rangle \langle y| \otimes \hat{D}_1 (t_f)\otimes \hat{D}_2 (0)\nonumber\\
 &
 +\int_{\mathcal{R}_2} \int_{\mathcal{R}_2} dx dy \hspace{1mm} \rho(x,y) |x\rangle \langle y| \otimes \hat{D}_1 (0) \otimes \hat{D}_2 (t_f)\nonumber\\
 &
+\int_{\mathcal{R}_3} \int_{\mathcal{R}_3} dx dy \hspace{1mm} \rho(x,y) |x\rangle \langle y| \otimes \hat{D}_1 (0) \otimes \hat{D}_2 (0)
\end{align}
where $\hat{D}_1$ and $\hat{D}_2$ stand for the first and second measurement spin apparatus, and  $\hat{D}_1 (0)$ represents the apparatus in its initial state 
(when it has not interacted with the spin) and $\hat{D}_1 (t_f)$ is the apparatus after the spin measurement has taken place. Therefore, given a quantum state
 described by a wavefunction and a spin, measuring its spin with two detectors spatially separated leads to a  truncation of the density matrix  
 (``collapse of the wavefunction'' either in one of the detectors or in the outer region). 
Although the density matrix  formally interacts with both apparatuses, only one apparatus produces a result.

In the considered process, the spin and the position degrees of freedom become entangled (because of the interaction (\ref{inttwoapparatuses})), 
and afterwards a measurement of the spin is carried out. This provokes a ``collapse of the wavefunction'' in both spin and position space. Note that a 
very similar situation was considered in the previous section, where a measurement of an entangled spin of two parties allows for inferring the state of both spins. 
This is translated here in the sense that the knowledge of the click (or not click) of the apparatus ascertains us to say that the system is localized (or not localized) there.

\section*{Conclusions}
The Curie-Weiss model describes a projective measurement of a spin as a physical interaction between the tested spin and a magnetic memory. 
Here we have applied  this model to the measurement of an entangled state, and that of a single spin by two spatially separated apparatuses. 
By using previous results in \cite{OpusABN}, we have straightforwardly obtained the final state of the spin and the apparatuses. The corresponding statistics 
coincide with those obtained by directly applying the measurement postulates, hence giving further support to the validity of the CW-model and the postulates alike. 

The present work complements a previous study on the simultaneous measurement of non-commuting observables \cite{marti}.
A natural complementary problem appears in the Stern-Gerlach experiment, where a position measurement is carried out in order to infer the spin.

\emph{Acknowledglements}
We thank Marian Kupczynski for correspondence.
This work was carried out during the master thesis research of M. P.-L. at the University of Amsterdam;
he thanks the members of the Institute for Theoretical Physics for hospitality.
M. P.-L. also acknowledges support from La Caixa Foundation.


\end{document}